\documentclass[12pt]{article}
\usepackage[utf8]{inputenc}
\usepackage{url}
\usepackage{amsmath}
\usepackage{amssymb}

\begin{document}
\begin{center}\Large{\bf Some remarks on the early history of the
    Albert Einstein Institute.}\\ \normalsize Hubert
  Goenner\\ Institute for Theoretical Physics\\ University
  G\"ottingen\\ Friedrich-Hund-Platz 1\\37077  G\"ottingen\end{center} 

\section{Introduction}
After the re-unification of Germany in 1990, reorganizations on all
levels of scientific activity took place, mainly in the former German
Democratic Republic (GDR). The universities there and the institutions
for research were adapted to the  existing structures in West Germany
by planning ``unsophisticated and conservative imitations of western
institutions'' as the then chairman of the German ``Science Council''
critically wrote (\cite{Simon1991}, p. 23). As a first step, the
Science Council initiated an evaluation of universities and research
institutes of GDR in June 1991 under the chairmanship of H. Gabriel,
Berlin. The early consultation of the Ministers for Research and
Science of the five new States, representing the former GDR, with the
president of Max Planck Society, Hans F. Zacher  (1928-2015), in
Munich from 9  to 10 December 1990 shows, that from the beginning the
Max Planck Society was a prominent one among many players
(\cite{MPG1989}. Part I, p. 371). Its aim was to reach the same
influence in the new States of the East as it already had in the West. \\ 

In GDR, research on relativistic theories of gravitation had been
carried through both at universities like the University of Jena and
in institutions of the Academy of Science of GDR like the
Zentralinstitut für Astrophysik (ZIAP) which incorporated
astrophysical institutes and astronomical observatories - among them
the Einstein-Laboratory. As I spent a sabbatical in winter 1990/91 at
the Technical University Berlin, I had contact with both, groups in
West-Berlin and at ZIAP, where I met Dr. H.-H. von Borzeskowski of the
Einstein-Laboratory in Potsdam-Babelsberg in November and attended a
meeting there on 8 December 1990. On 17 January 1991 I gave a lecture
at ZIAP; a meeting with Prof. D.-E. Liebscher (Einstein-Laboratory)
occured on 6 February 1991. Such activities must be seen in connection
with preparations for 
the memorandum with F. Hehl to be introduced below. The chairman of
the German Astronomical Society at the time, Wolfgang Hillebrandt,
also visited ZIAP on 4 February 1991 in order to be able to give a
recommendation on its future\footnote{This reference was pointed out
  to me by G. Schäfer, Jena} \cite{Hille2013}. He 
simultaneously was a scientific member of the Max Planck Institute
(MPI) for Physics and Astrophysics in Munich. According to him: ``Of
course, the question was raised whether, and if yes, in what way
Max Planck Society and its institutes would be able to help''. But for
financial reasons: ``[..] in the end, we could offer our colleagues at
ZIAP only our moral support [..]'' (\cite{Hille2013}, p. 135).\\

 
In April 1991, another initiative came to life: the ``Arbeitsstelle
Albert Einstein in Berlin''\footnote{This was a research project of J. Renn
at the Max Planck Institute for Human Development under its director
Wolfgang Edelstein, financed by the Senate of Berlin for 5 years.} 
\cite{ArBe91}. It may be seen as one of the precursors of the Max Planck
Institute for the History of Science, founded in 1994, which absorbed
a philosopher, originally beloging to Treder's Einstein-Laboratory. \\   

  In July 1991, the German Council of Science recommended that the
  Zentralinstitut für Astrophysik, and particularly, H.-J. Treder's
  Einstein-Laboratory for Theoretical Physics in Potsdam, both of
  which were involved in research on general relativity, should be
  discontinued: ``In its present size and constitution, the
  Einstein-Laboratory does not offer the conditions for a sufficiently
  broad contribution to the complex problems of gravitational theory,
  modern cosmology and the unification of general relativity with
  quantum theory [..].''(\cite{Wissrat1991}, p. 86.) 

Somewhat earlier, on 8 March 1991, the Max Planck-Institute (MPI)
for Physics and Astrophysics in Munich had been split up into three
independent institutions, the MPI for Physics in Munich, the MPI for
Astrophysics and the MPI for extraterrestrial Physics both in Garching
with the last two originally established as subdivisions already in
1963. The working group ``gravitational theory'' under its director
Jürgen Ehlers (1929-2008) installed in 1971 continued within the MPI
for Astrophysics. Ehlers had an internationally recognized scientific
stature and strove to control the field in Germany. As we will see
below,  astrophysics in GDR became directly related to the foundation
of a new Max Planck Institute for gravitation.\footnote{I must rely
  mostly on documents privately held, because the retention period for
  the intervall 1990 to 1995 in the archive of Max Planck Society is
  still valid. Nevertheless, I could look at a few documents which
  strengthened and complemented my point of view.} \\  

\section{Preludes to the foundation of the Albert Einstein Institute} 
\subsection{Suggestion of an International Einstein Center}
 On 8 February, 1991, four months after the German ``reunification'' and
 {\em prior} to the recommendation of the Council of Science mentioned above,  Friedrich Hehl, Cologne and Hubert Goenner, Göttingen, formulated a 
Memorandum pleading for the foundation of an {\bf International Einstein
Center} in Potsdam/Caputh. After a description of the historical
situation, and the low importance given to research on general
relativity in Germany - with two leading scientists nearing retirement
- it was stated there:
 \begin{quote} ``[..] Quite certainly, no 'relativist' will be appointed
   to the full professorships mentioned above after retirement of the
   present incumbents. As seen from the international standard of
   competition in a fundamental branch of modern physics, for junior
   researchers this situation is, consequently, rather discouraging in
   terms of job openings etc. A closing down of the
   Einstein-Laboratory, without substitution, would appear
   irresponsible under such circumstances. [..] {\em In summing up, we
     suggest the creation of the International Einstein Center (ICE)
     in which theoretical physics research in the field of relativity
     and gravitation in relationship with elementary particle physics
     is cultivated}.''\end{quote} Special emphasis was put on the {\em
   international} character of ICE, in the sense that not just one
 country alone should become involved in the foundation.\\ Also
 Einstein's summer home in Caputh, in GDR belonging to the
 Einstein-Laboratory, was included in this plan to hold up Einstein's
heritage in Germany.\footnote{Had we known about the complicated
  problems with ownership concerning Einstein's summer house arising only
  later, we certainly would have abstained from such a consideration.}
The summer home now is administered by the ``Einstein Forum''
established in 1993 by the State of Brandenburg with the participation
of members of the Hebrew University, Jerusalem, and the Swedish Academy
of Sciences.\\

This memorandum was submitted to the secretary of the German Council
of Science and Humanities (Wissenschaftsrat) on 11 February 1991 and later
announced in ``Physikalische Blätter'', a journal related to German
Physical Society \cite{GoeHe1991}. In it, the hope was expessed that
German junior relativists would  finally find a secure opportunity for
pursuing their research. Also, the Minister of Science of the new 
state of Brandenburg, Hinrich Enderlein was informed and expressed his interest  by earmarking an amount in the budget for this purpose
\cite{Ender1992}. Yet, a blocking notice was marked for the case of
`` Third party conception and trusteeship'' \cite{Brand1991}. The then
German Minister for Research and Technology, Heinz Riesenhuber, also
was approached. Although open-minded with regard to our suggestion, he
wanted to wait for a decision by the Council of Science
\cite{Ries91}. I also sent a copy of the memorandum to our colleague
Jürgen Ehlers at the MPI for Physics and Astrophysics, Munich on 28
February 1991 whom I had orally informed already during a previous
encounter at a F.E.S.T-seminar in Heidelberg. \\       
\subsection{A possible cooperation between Germany and Israel?}
Originally, our intention was to establish a joint German-Israeli
research institute and to ask the German-Israeli Foundation for
Scientific Research \& Development (GIF) to take part besides the
German Federal Ministry for Research \& Technology and the State of
Brandenburg. The plan won the approval of the well-known Israeli
theoretical physicist Yuval Ne’eman (1925-2006) \cite{Neem1991}, at
the time Minister for Science as well as Minister for Energy and
Infrastructure.\footnote{Y. Ne’eman co-discovered SU(3)-symmetry in
  particle physics. He was president of Tel Aviv University
  (1971-1975) and founder of the Israel Space Agency in 1983.} This
explains why in the {\em committee of trustees} we suggested, two
theoretical physicists from Israel appeared: Y. Ne'eman (Tel-Aviv) and
N. Rosen (Haifa). In addition, H. Fritzsch (Munich), 
Abdus Salam (Trieste/London), E. Schmutzer (Jena), and J.A. Wheeler
(Princeton) were named. J. A. Wheeler, due to his inability ``to
estimate the political factors from this distance, and due to existing
heavy commitments'' did not want to become a trustee\footnote{It is
  unclear whether Wheeler refered to an eventual political involvement of
  H.-J. Treder (1928-2006), a doctoral student of A. Papapetrou
  (1907-1997) \cite{Hoff2016}. Treder was member of the communist
  party in West Berlin and later had good relations to members of the
  central committee of the ruling party in GDR (SED). However, he was
  not politically active himself.} \cite{Wheeler1991}. Also, the board
of GIF communicated ``with regret'' that, due to its bylaws GIF could
not permanently support 
such an Einstein Center \cite{Bar91}. As it turned out, the State of
Israel was interested neither in a German nor in an international
Institute, but only in an eventual establishment of {\em two} Einstein
Centers, one in Jerusalem, the other in Germany, possibly under the
auspices of the MINERVA foundation\footnote{In 1964, the Minerva Foundation was
  established as a subsidiary of the Max Planck Society; ever since
  then it is financed by the German Federal Ministry for Education and
  Research.}. However, as Israel was short of the money needed, there
was no hope in its involvement.\footnote{Copies of the letters by GIF
  also went to the Federal Minister Riesenhuber.} As a way out for
such a case, in the memorandum by Hehl and Goenner a possible
engagement of ``UNESCO and other organizations'' had been
mentioned. Perhaps, we were overly optimistic: this unconventional
suggestion was not taken up by any of the relevant organizations in
Germany. At the time of re-unification, politics including science
policy, strove to arrange the incorporation of the former GDR with the
minimal number of new ideas.\footnote{To give another example: all attempts
  for working out a constitution for re-united Germany replacing the
  temporary ``Grundgesetz'' were stalled by the ruling political parties.}\\   

Further international support for our initiative came from a number
of well-known colleagues like Einstein’s former assistant
P. G. Bergmann, Syracuse, B. Bertotti, Pavia, and Allan Held, Bern,
editor of the journal of relativity and gravitation. P. Bergmann and
V. de Sabbata, Bologna, started an initiative of their own for the
maintenace of Treder’s Einstein-Laboratory as well as Wolfgang Edelstein and Peter Damerow (1939-2011) from the MPI for Human Development \cite{Edel91}. \\

\subsection{How the Max Planck Society took over}
Fully aware of our initiative and of the missing institutional
background in terms of funds and positions on the side of the
initiators, in July 1991 the German Council of Science and Humanities 
suggested the {\em foundation of an Albert-Einstein-Institute for
  Gravitational Physics in the region
  Berlin-Potsdam}: \begin{quote}``Institutions from abroad with
  corresponding competences should take part in the foundation. The
  Council of Science asks the Max Planck Society to take over the lead
  management for the appointment of a working group which makes
  suggestions for a possible organizing institution and an adequate
  infrastructure.'' (\cite{Wissrat1991}, p. 88.) \end{quote} At
this point, in 1991, MaxPlanck Society was still far from assuming the
trusteeship for the suggested institute. In fact, according to its
president Hans Zacher: ``[..] this commission merely will
address the task to build a scientific conception for the suggested
[by the Council of Science] Institute for Gravitational Physics. The
Memorandum resulting from these deliberations will be the basis for
the State of Brandenburg to appoint the actual founding commission for
the suggested institute [..]''\cite{Zacher1991}.\\  

This way of proceeding was supported by the then president
of German Physical Society (DPG), Th. Mayer-Kuckuk (1927-2014)
\cite{MayKu1991}. To him and to the president of Max Planck Society,
F. W. Hehl and H. Goenner then expressed their full support for an
initiative in this direction by the Max Planck Society, and also their
wish to take part in a preparatory group \cite{GoeHe1991b},
\cite{GoeHe1991c}. The Fachverband Gravitation and Relativity through 
its chairman, G. Schäfer, did not support our memorandum but instead
the recommendation of the German Council of Science which had included
essential elements of the memorandum by Hehl \& Goenner. H. Schäfer at
the time belonged to the group of J. Ehlers \cite{Schaef1991}. The
suggested working group was {\em not} set up by the State of
Brandenburg as claimed by Zacher, but due to a letter by the
Brandenburg minister that he should take it in his hands, H. Zacher
delegated it to J. Ehlers who became its chairman; it included the
acting director of the MPI for Astrophysics, W. Hillebrandt, and three
expert colleagues from universities in Jena, Paris and Zürich. To the 
German physics community, in particular to the Fachverband Gravitation
and Relativity, neither the members of this group nor its proceedings
were communicated. For discussions concerning the funding of the new
institute, Ehlers met with people from the ministry in Brandenburg at
the end of October 1991. The financial means to be contributed by the
State of Brandenburg were insufficient. A report from Ehlers envisaged
at first for the end of 1991 and then for the time after a meeting of
his preparatory group in January 1992 did not materialize for the
public although it was ready since the end of 1991. The internal
discussions in the Max Planck Society have been lengthy and less
than unanimous.\\  
During the summer of 1992 a rumor, triggered by the retirement of the
director of the MPI for Astrophysics in Munich, R. Kippenhahn, came up that this
institute which also housed the gravitational group directed by
J. Ehlers, might become closed down \cite{PhysBl1992}. In fact, four
alternatives were discussed within Max Planck Society: (1) to join the
MPI for Astrophysics to the MPI for Extraterrestrial Physics; (2) to
distribute the activities of the Max Planck Society in the field of
Astrophysic among the existing institutes in Heidelberg, Bonn and
Garching; (3) an amalgamation with ZIAP or part of it (suggested by
G. Haerendel of the MPI for Extraterrestrial Physics), and (4) the
maintaining the institute as it existed \cite{Trumpera16}. In his
remarks concerning the foundation of the Max Planck Institute for
Gravitational Physics, W. Hillebrandt reports a suggestion, told him
by H. Zacher that the MPI for Astrophysics should be merged with
theoretical groups of ZIAP and moved to Potsdam, and that he refused
to concur (\cite{Hille2013}, p. 135). Perhaps, this was the expression
of a trend of the time, i.e., to reduce funds for science in the West
of BRD in favour of new structures in the East: ``In the landscape of
research, buildup of the East by cutback in the West cannot always be
excluded.''\footnote{``Aufbau Ost durch Abbau West kann auch in der
  Forschungslandschaft nicht immer ausgeschlossen werden''}\cite{Cat93}. \\    

At the meeting of Max Planck Society on 8 March 1991, it was also
decided that, as a first measure, G. Neugebauer of the University Jena
should be asked by H. Zacher to establish in Jena one of the 27
working groups supported by the Society in 1991 and 1992 for a
duration of five years. Its begin was set to 1 January 1992
(\cite{MPG1989},  part I, p. 375).\footnote{The original decision for
  the establishment of such working groups and the code of practice
  dates from 5 November 1990 (\cite{MPG1989}, part I, p. 370).} In the
tradition of Max Planck Society, Neugebauer, in principle, was free to
choose the three other members. His partner institute was the MPI for
Astrophysics in Garching. The suggestion for this group in Jena had come
from J. Ehlers and on his recommendation Gerhard Schäfer from the
group in Garching joined Neugebauer's group as the fourth member in March 1992. \\    

It was not before 19 July 1993 that the ``Memorandum on the
founding of an Albert Einstein Institute für Gravitationsphysik''
finally was issued by the {\em Working Group} set up by J. Ehlers and 
distributed by him\footnote{He had written the memorandum and its
  support was secured through a circulation procedure.}. The memorandum emphasized: \begin{quote} ``What is
  missing is an institute where researchers from Germany and abroad
  can collaborate for reasonable periods of time. An Einstein
  Institute could serve this purpose and thus stimulate also both
  research and teaching at universities. Universities cannot play this
  role: Positions are not available, high-level teaching requires a
  minimal number of people with small teaching obligations working in
  close contact with each other and with guests from abroad. [..] The
  research should mainly concentrate on {\em basic physics} not on astrophysical
  phenomenology.''\end{quote}  Thus, similar to what had been
formulated during the 3rd Hochschulreform of GDR in 1968
\cite{Laitko1997}, the 
intention was to clearly distribute tasks between low-level teaching
at universities and research-oriented high-level teaching in close 
cooperation with Max Planck Institutes. It is surprising that the
university professors on the ``Working Group'' supported such a
formulation: They were {\em obliged} to teach but knew that teaching
was {\em voluntary} for members of a Max Planck Institute. In
addition, the memorandum  accepted the lack of positions for
relativity research in Germany as unalterable. ``It may be necessary
to obtain leading scientists from abroad to direct such an
institute.''\footnote{As it turned out later, from Germany only junior
  scientists belonging to Ehler’s group inside the Max Planck Society
  would obtain permanent positions in the new institute.} In the
memorandum, it was made very clear that only ``the  trusteeship of Max
Planck Society'' could secure the plan: \begin{quote} ``The committee
  feels that the optimal solution,   and perhaps the only one which
  would guarantee that a ``center of excellence'' with a long-time
  perspective could actually be formed, would be that the
  Max-Planck-Gesellschaft founds such an institute, to be named
  Max-Planck-Institut für Gravitationsphysik (Einstein-Institut). [..]
  In accordance with this memorandum, the committee recommends that
  Jürgen Ehlers apply for the founding of a Max-Planck-Institut für
  Gravitationsphysik.'' \end{quote} Unlike what had been intended by
the memorandum of Hehl and Goenner and by the statement of the German
Council of Science, the new institute would rest on a purely {\em
  national} basis but with an international personnel. This was an
early expression of the present policy of Max Planck
Society: \begin{quote} ``Against the background 
  of the high degree of internationalization of MaxPlanck Society -
  as seen on an international scale, - and of the resulting high scientific
  performance capacity, [..] Max Planck Society further continuously
  boosts the process of
  internationalization.''\cite{MPG2016} \end{quote} Together with its
memorandum, Ehlers sent around invitations for comments from
representatives of the international and German community of
relativists on the occasion of an announced ``Symposium on
Developments and Trends in Gravitational Physics'' to be held on
Sept. 20-21, 1993 in Munich. During the symposium, besides the
scientific lectures,\footnote{Of the thirteen scientific talks, only 
  three were given by speakers from the German relativity community.}
the eventual structure of the planned institute was discussed (types
and number of positions) as well as possible candidates for
directorship. Max Planck Society had followed the recommendation by
introducing a second ``Scientific Organization Committee'' with
G. Wegner, MPI for Polymer Research as its chairman and 6 other
directors of Max Planck Institutes plus 3 university professors as 
members. J. Ehlers and H. Walther, MPI for Quantum Optics, at the time
vice-president of Max Planck Society were present as guests.\\

In view of its past difficulties with Max Planck Society, the MPI for
Astrophysics might not have been unhappy to loose Ehler’s
department. At the time, it seems not to have played a big role in the 
institute: in the three yearly reports from 1991 to 1993, no
topic from Ehler’s group is noted among the key activities of the MPI
for Astrophysics in Garching\footnote{This is somewhat surprising because
  during these years Ehlers' group and his guests contributed 20\% of
  all publications from the MPI for Astrophysics. Cf. \cite{MPGJV90}.}
(\cite{MPGJ91}, p. 303-313, \cite{MPGJ92}, p. 311-319, \cite{MPGJ93},
p. 349-359). \\  

On 9 June 1994, during its meeting in Göttingen, the Max Planck
Society finally decided to establish an Institute for Gravitational
Physics ``in the region of Potsdam'' with J. Ehlers, a scientific member of
Max Planck Society, as one of the envisaged three
directors\footnote{Bernhard F. Schutz from the University of Cardiff
  was also considered as a possible director.} (\cite{MPG1989}, part I,
p. 416). 16 permanent positions and the same number for visiting
scientists were planned.  In its yearly report for 1994, very
fittingly, now the MPI for Astrophysics in Garching highlighted
Ehler’s group under ``Global dynamics of self-gravitating matter''
(\cite{MPGJ94} p. 343-346). 
\section{Opening and growth of the institute}
Almost one year later, the new Max Planck Institute for Gravitational
Physics \linebreak (``Albert Einstein Institute'') opened in Potsdam on
1 April 1995.  (\cite{MPG1989},
p. 426). The original spin doctors for the foundation of such an
institute, i.e., F. W. Hehl and H. Goenner, were not
invited.\footnote{This is in conformity with the present formulation
  of the institute's history: ``Its establishment was an initiative of
  its founding director, Jürgen Ehlers (1929-2008)'' \cite{AEI2016}.}
The institute began with two departments headed by J. Ehlers (``Physical
Foundations and mathematical methods of general relativity'') and
Bernard F. Schutz (``Gravitational theories oriented toward
observations'') who began to work in his position in June 1995. The
third department directed by Hermann Nicolai, Hamburg, started on 1
March 1997. Originally, it had been planned as a working area for
``Relativity- and quantum theory'' (\cite{MPGJ95}, p. 423). At the end
of 1998, J. Ehlers already retired: he had been director for only 3
and a half years, an exceptional situation for a newly created Max
Planck Institute. Since 1999, the Albert Einstein Institute has moved
to its new building in Golm near Potsdam. \\     

Unfortunately, by its construction the structure of the 
Albert Einstein Institute showed that general relativity was
considered as an appendix to either astrophysics and elementary
particle physics, or to mathematics. The institute established a
leading international role in research: now, it seems to be the
largest institute in the world for research on relativistic
gravitation. While countless guests from abroad were welcomed on
German taxpayers’ money, the job-situation in Germany for relativists
was not improved by the Albert Einstein Institute (AEI). Together with
the University of Potsdam, programs leading to a PhD were offered. The
steady number of about 14 PhDs produced per decade at German
universities in the field of general relativity, cosmology and
relativistic astrophysics during the three decades from the 1960s to
the 1980s, in the first years after 1995 remained uninfluenced by the
new Max Planck Institute for Gravitational Physics. This situation has
changed dramatically, however. At present the AEI in Golm trains 19
PhD students, i.e., probably more in the field of gravitation than any German
university does.\\  

In 2002, to the Albert Einstein Institute a fourth section on
experimental gravitation in Hannover (measurement of gravitational
waves by interferometry, data analysis) with its director Karsten Danzmann
has been added. In 2007 Bruce Allen became a further director
concerned with gravitational waves. The institute is rooted both in
the MPI for Quantum Optics, Munich, where the first interferometers
were built, and in the Institute for Atom and Molecule Physics (AMP)
of the University of Hannover.\footnote{Since 1 January 1994 the MPI
  for Quantum Optics had a branch office in Hannover.} It runs the
detector GEO 600 for gravitational waves with arms of 600 meter,
together with the universities of Glasgow and Cardiff. The institute
has contributed importantly to the recent great success of the direct
observation of a gravitational wave by the LIGO-group in the United
States. At present, AEI-Hannover has attracted 44 PhD students.   

\section{Conclusion}
How has the landscape for research on general relativity and other
relativistic theories of gravitation changed since the opening of the
Max Planck Institute for Gravitational Physics? The most important new
feature is that research in gravitation in Germany now has a fixed point
- independent of the appointment of full professors at universities,
more or less at haphazard with regard to a rational policy of keeping
research on gravitation going. The total number of 
PhD-students at the MPI for Gravitational Physics seems to
show that, at present, at this single institution more PhDs are
trained on gravitational topics than by professors in all German
universities together. The situation is more complicated, however. On
the one hand, the subjects of research at the Max Planck Institute for
Gravitational Physics in Golm - outside the ``Astrophysical and
Cosmological Relativity Division'' - have broadened 
to the extent that only a small part of these PhDs will be in general
relativity proper and quantum gravity. And on the other hand, the
great majority of the PhD students are coming from abroad - very much
in the spirit of the cultural politics of the Ministry of Foreign
Affairs. One wonders how this situation will continue if less and less
professorships connected to research in relativistic theories of
gravitation at German universities are filled. The trend seems to go
toward applied gravitational research in astrophysics and
astronomy.\footnote{By the creation of graduate schools and graduate
  colleges including the words ``gravitation'' or ``cosmology'' in
  their title nothing more than the {\em financing} of doctoral
  degrees is reorganized.} \\The genesis of the Max Planck Institute for
Gravitational Physics as described here, hopefully, shows that the
history of the founding of the institute is not as simplistic as claimed: that
it came to life because J. Ehlers had the idea and realized his
``lifetime dream of a Max Planck Institute for Gravitational
Physics''\footnote{It looks rather that someone with good relations to
president H. Zacher had ``jumped the band waggon'' and then worked
hard and succeeded to direct the ``waggon'' to where he wanted it to go.}
(\cite{Hille2013}, p. 137). In contrast, it is an example for the
observation that scientists: ``[..] are strategists, choosing the most opportune
moment, engaging in potentially fruitful collaborations, evaluating
and grasping opportunities, and rushing to credited information. Their
political ability is invested in the heart of doing science.''
(\cite{LaWoo1986}, p. 213). A condensed formulation is given by Kohler
for whom scientists: ``build careers by occupying a number of
strategic positions on an incessantly changing market''
(\cite{Kohler1989}, p. 166).\\ 
  
\section{Acknowledgment}
My gratitude goes to Friedrich W. Hehl, Cologne, whose enterprise
played a crucial role for our initiative. He shared his memory with
me, allowed me to use his correspondence, and carefully read various
previous versions of the manuscript. 
I also thank 
D. Hoffmann for discussions and pointing out references. I am
grateful to the director of the archive of Max Planck Society, Dr. Kristina
Starkloff, for allowing me to look at some documents fitting to this research.   

\end{document}